\documentstyle[12pt,aaspp4,tighten]{article} 

\def\pagebreak{\vfill\eject}

\def\sun{\mbox{$_\odot$}}

\def\deg{{$^\circ$}}
\def\half{{\leavevmode\kern.1em\raise.5ex
\hbox{\the\scriptfont0 1}\kern-.1em /
\kern-.15em\lower.25ex\hbox{\the\scriptfont02}}} 
\def\gtsim{\lower.5ex\hbox{$\buildrel > \over\sim$}}
\def\ltsim{\lower.5ex\hbox{$\buildrel < \over\sim$}}
\def\sun{\mbox{$_\odot$}}
\def\kms{~km~s$^{-1}$}

\def\h{H$_2$}

\def\water{H$_2$O}

\def\mjybkms{~mJy~beam$^{-1}$ km~s$^{-1}$}
\def\mjyb{~mJy~beam$^{-1}$}

\def\dec#1#2#3{$#1^\circ#2'#3''$}

\slugcomment{ApJ, in press (2004, v 601, Feb 1 issue)}

\begin{document}

\title{The Nature of the Massive Young Stars in W75~N}

\author{D.~S.~Shepherd\altaffilmark{1},
S.~E.~Kurtz\altaffilmark{2}, 
L.~Testi\altaffilmark{3}}

\vspace{-3mm}
\altaffiltext{1}{National Radio Astronomy Observatory, P.O. Box 0,
Socorro, NM 87801}
\altaffiltext{2}{Centro de Radioastronom\'\i a y Astrof\'\i sica,
Universidad Nacional Aut\'onoma de M\'exico, Apdo. Postal 3-72, C.P.
58089, Morelia, Mich. Mexico}
\altaffiltext{3}{INAF -- Osservatorio Astrofisco di Arcetri, Largo Enrico
Fermi 5, I-50125 Firenze}

\rightskip=\leftskip 
\vspace{-3mm}

\begin{abstract}

We have observed the W75~N massive star forming region in SiO(J=2--1
\& J=1--0) at $3'' - 5''$ resolution and in 6~cm, 2~cm, and 7~mm
continuum emission at $1.4'' - 0.2''$ resolution.  The abundance ratio
of [SiO]/[\h] $\sim 5-7 \times 10^{-11}$ which is typical for what is
expected in the ambient component of molecular clouds with active star
formation.  The SiO morphology is diffuse and centered on the
positions of the ultracompact HII regions - no collimated, neutral jet
was discovered.  The ionized gas surrounding the protostars have
emission measures ranging from $1-15 \times 10^6$~pc~cm$^{-6}$,
densities from $0.4-5 \times 10^4$~cm$^{-3}$, and derived spectral
types of the central ionizing stars ranging from B0.5 to B2.  Most of
the detected sources have spectral indicies which suggest optically
thin to moderately optically thick HII regions produced by a central
ionizing star.  The spread in ages between the oldest and youngest
early-B protostars in the W75~N cluster is $0.1-5 \times
10^6$ years.  This evolutionary timescale for W75~N is consistent with
that found for early-B stars born in clusters forming more massive
stars ($M_\star > 25$M\sun).

\end{abstract}


\clearpage
\section{INTRODUCTION}

Molecular outflows from young, early-B protostars have many
characteristics in common with those from lower mass young stellar
objects (YSOs) while the HII regions produced by the central stars
often look similar to those produced by O stars.  For example, the
outflow momentum and the mass of circumstellar material both scale
with the bolometric luminosity of the driving source (e.g. Levreault
1988; Cabrit \& Bertout 1992; Rodr\'{\i}guez et al. 1996; Shepherd \&
Churchwell 1996; Chandler \& Richer 2000).  Some mid- to early-B YSOs
have ionized or molecular jets that are well-collimated close to the
YSO (e.g. IRAS 20126+4014: Hofner et al. 1999; Ceph A HW2: Torrelles
et al. 1993; Rodr\'{\i}guez et al. 1994; Garay et al. 1996); at least
one source, HH~80--81, has a well-collimated, parsec-scale ionized jet
that appears to be a scaled version of a Herbig-Haro jet from a
low-mass YSO (Mart\'{\i}, Rodr\'{\i}guez, \& Reipurth 1993; Heathcote,
Reipurth, \& Raga 1998).  Yet despite these similarities, recent
observations have shown that the characteristics of early-B star
outflows and disks may also be diverging from their low-mass
counterparts.  In particular, molecular outflows from early-B stars
tend to be less collimated than those from low-mass YSO even when
there is a well-collimated, ionized jet (e.g. HH~80--81: Yamashita et
al. 1989; IRAS 20126+4014: Shepherd et al. 2000), and some outflows
show no evidence for a collimated jet (G192.16--3.82: Shepherd,
Claussen, \& Kurtz 2001), instead sporting a classic ultracompact (UC)
HII region at the protostellar position.

One early-B star cluster with outflows that may exhibit some
differences from low-mass flows is W75~N: a massive star forming
region with an integrated IRAS luminosity of $1.4 \times 10^5$~L\sun\
forming mid- to early-B stars (Haschick et al. 1981; Hunter et
al. 1994; Torrelles et al. 1997).  

At the heart of the W75~N outflows is a cluster of four UC HII regions
embedded in a millimeter core W75~N:MM~1\footnote{Names of millimeter
cores are shortened to MM~1-5 for the remainder of this paper}.
Haschick et al. (1981) identified three regions of ionized gas in
W75~N at a resolution of $\sim 1.5''$: W75~N (A), W75~N (B), and W75~N
(C).  Hunter et al. (1994) later resolved W75~N (B) with $\sim 0.5''$
resolution into three regions: Ba, Bb, and Bc.  Torrelles et
al. (1997) then imaged W75~N (B) at $\sim 0.1''$ resolution, and
detected Ba and Bb (which they called VLA~1 \& VLA~3), along with
another weaker, and more compact HII region, VLA~2.  Within a $10''$
radius of MM~1 are three, compact millimeter cores (MM~2-4).  None of
these sources have discernible near-infrared counterparts although
there is substantial near-infrared reflection nebulosity in the region
(see, e.g., Figs 1, 5, \& 10 from Shepherd, Testi, \& Stark 2003,
hereafter STS03).  Mid-infrared emission at
12.5$\mu$m has been detected in the vicinity of the UC HII regions
however it is unclear which source(s) are producing the emission
(Persi et al. 2003).  An extended millimeter core (MM~5) is located
roughly $30''$ north of MM~1 and has an associated reflection nebula
(W75~N A) and central star visible in the infrared.  

Multiple outflows have been identified originating from the cluster of
UC HII regions and millimeter cores with a total flow mass greater
than 250~M\sun\ (Fischer et al. 1985; Hunter et al. 1994; Davis et
al. 1998a, 1998b, Ridge \& Moore 2001; Shepherd 2001; STS03, Torrelles
et al. 2003).  Davis et al. (1998a,b) suggest the outflow is driven by
a powerful, well-collimated jet while STS03 find no evidence for a
jet.  But is there an underlying, undetected neutral jet driving the
flow?  And what are the properties of the HII regions and are they
consistent with what is expected for ionized gas around early-B
zero-age-main-sequence (ZAMS) stars?  To answer these questions we
have observed W75~N in SiO(J=2--1 \& J=1--0) line emission to search
for evidence of a neutral jet and in centimeter \& 7~mm continuum
emission to obtain a better understanding of the nature of the
powering sources.

\section{OBSERVATIONS}

\subsection{Owens Valley Radio Observatory}
Observations in SiO(v=0, J=2--1) and SiO(v=1, J=2--1) were made with
the Owens Valley Radio Observatory (OVRO) millimeter-wave array of six
10.4~m telescopes between 1999 May 26 and 1999 November 12.  The 64
channel spectral bandpass was centered on the local standard of rest
velocity ($v_{LSR}$) of 10{\kms} with a spectral resolution of
1.726{\kms} for SiO(v=0) and 1.738{\kms} for SiO(v=1).
Gain calibration used the quasar BL~Lac while observations of Neptune
and/or Uranus provided the flux density calibration scale with an
estimated uncertainty of $\sim 20$\%.  Calibration was carried out
using the Caltech MMA data reduction package (Scoville et al. 1993).
Images were produced and analyzed using the MIRIAD software package
(Sault, Teuben, \& Wright 1995).  SiO(v=1, J=2--1) was not detected.
SiO(v=0, J=2--1) was detected and images were made with and without a
Gaussian taper of $3''$ FWHM to optimize sensitivity to extended
structure and more compact features, respectively.  A summary of the
observational parameters is presented in Table 1.  The largest angular
scale that can be accurately imaged, $\theta_{LAS}$, is $\sim 20''$.

\begin{table}[h]
\caption[]{Observational Summary}
\smallskip
\begin{tabular}{|c|c|c|c|c|c|c|}
\hline
      & Rest      & Beam & Beam &       & Peak$^\dagger$       
      & Total$^\dagger$  \\
Image & Freq & FWHM & P.A. & RMS     & Flux Density
      & Flux Density \\
      & (GHz)     & (arcsec) & (deg) & (mJy/beam) 
      & (mJy/beam) & (mJy) \\
\hline
\hline
6~cm continuum  & ~4.88 & $1.36 \times 1.12$ & 29.2
       & ~~0.11 & ~~4.5  & ~~128.5 \\
2~cm continuum  & 14.96 & $0.46 \times 0.38$ & 34.5
       & ~~0.23 & ~~4.4  & ~~~~112.7 \\
7~mm continuum  & 43.34 & $0.27 \times 0.20$ & 89.4
       & ~~0.31 & ~~5.4  & ~~~~8.3 \\
SiO(v=0, J=1--0) & 43.42 & $0.40 \times 0.36$ &  ~65.8
       & ~~4.0~  & ~~14.0~  & ~~~14.0 \\
SiO(v=0, J=2--1) & 86.85 & $5.35 \times 4.15$ & --60.3
       & ~45.0~  & ~~640.0~~~  & 42,300.~ \\
SiO(v=0, J=2--1)$^{\dagger\dagger}$  
                & 86.85 & $3.17 \times 2.56$ & --61.5
       & ~39.0~  & ~~330.0~~~  & 17,700.~ \\
SiO(v=1, J=2--1) & 86.24 & $4.87 \times 3.83$ & --68.3
       & ~45.0~  & \nodata  & \nodata \\
\hline
\end{tabular}

\vspace{.1in}
~~{\small $^\dagger$~~Flux densities measured in the primary beam
	corrected image.  Total flux density given is the combined
	value for all sources in the field.} 

~~{\small $^{\dagger\dagger}$~Higher resolution, the more extended emission,
          58\% of the total flux density, has been resolved out.} 

\end{table}

\subsection{Very Large Array}

Observations of 43.3399~GHz (7~mm) continuum emission were made with
the National Radio Astronomy Observatory's\footnote{The National Radio
Astronomy Observatory is a facility of the National Science Foundation
operated under cooperative agreement by Associated Universities, Inc.}
Very Large Array (VLA) in the ``C'' configuration on 2000 April 24 and
in the ``B'' configuration on 2001 March 22.  Baselines between 35~m
and 11.4~km could detect a largest angular emission scale,
$\theta_{LAS} \sim 18''$.  The total bandwidth was 200~MHz.  The
quasar 2012$+$464 was used as a phase calibrator and 3C286 was the
flux calibrator. The estimated uncertainty of the flux calibration is
$\sim 10$\%.  Calibration and imaging was performed using the AIPS$++$
data reduction package.  The data were imaged using natural $uv$
weighting and the image was deconvolved with a CLEAN-based algorithm.

Observations of 4.8851~GHz (6~cm) and 14.9649~GHz (2~cm) continuum
emission were made with the VLA in the ``B'' configuration on 2001
March 22.  Baselines between 0.21~km and 11.4~km detected a largest
angular scale $\theta_{LAS} \sim 26''$ at 6~cm and $9''$ at 2~cm.  The
quasar 2012$+$464 was used as a phase calibrator, the quasars 3C286
\& 0410+769 were used as flux calibrators. The estimated uncertainty
of the flux calibration is $\sim 1$\% for both 2~cm and 6~cm.
Calibration and imaging was performed using the AIPS$++$ data
reduction package.  The data were imaged using robust $uv$ weighting. 
The 2~cm image was deconvolved with a CLEAN-based algorithm while the
6~cm image was deconvolved using both a standard CLEAN-based algorithm
and a multi-scale CLEAN algorithm.  The multi-scale image was CLEANed
with six component scale sizes with diameters of $0''$, $.6''$,
$1.2''$, $2''$, $4''$, \& $6''$.  Both 6~cm images gave essentially
the same flux however the image generated with multiple scales
provided the best sensitivity to extended emission in the HII region
W75~N~A while preserving the compact structure in the UC~HII regions
in W75~N~B.  In contrast, the 2~cm (and 7~mm) observations resolved
out most of the flux in W75~N~A to the point where it could not be
imaged properly and the standard CLEAN algorithm was the most
effective deconvolution method.

SiO(v=0, J=1--0) line emission was observed with VLA in the ``C''
configuration on 2000 April 24.  The observations
were made using 32 channels centered on $v_{LSR}$ with a spectral
resolution of 2.7{\kms} and a total bandwidth of 86.4{\kms}.  The
quasar 2012$+$464 was used as a phase calibrator and 3C286 was the
flux calibrator. The estimated uncertainty of the flux calibration is
$\sim 10$\%.  Baselines between 35~m and 3.4~km provided a synthesized
beam of approximately $0.4''$ over the $1'$ field of view.  The
largest angular scale that the VLA is sensitive to at this frequency,
$\theta_{LAS}$, is $\sim 18''$.  Calibration and imaging was performed
using the AIPS data reduction package.  The data were imaged using
robust $uv$ weighting and deconvolved with a CLEAN-based algorithm.
At $0.4''$ resolution, a single 14\mjyb\ peak ($3 \sigma$) in one
channel was recovered near the UC HII region positions.  Convolving
the $uv$ data with a Gaussian taper to yield a $3''$ beam resulted in
a peak emission of $\sim 30$\mjyb.  

\section{RESULTS}

\subsection{Continuum Emission}

The locations of the HII regions discussed in Section 1 are shown in
images of the ionized gas in 6~cm, 2~cm, and 7~mm continuum emission
(Fig. 1).  Peak and total flux densities are given in Table 2 and the
spectral energy distributions (SEDs) for each UC HII region are shown
in Fig. 2.
\small
\begin{table}[h]
\caption[]{Measured Flux Density (S$_{\nu}$) of Continuum Sources} 

\smallskip
\begin{tabular}{|l|c|c|c|c|c|c|c|c|}
\hline
&         &Total         &Peak          &Total         &Peak 
        &Total         &Peak          &   \\ 
&Position &S$_{6cm}$ &S$_{6cm}$ & S$_{2cm}$& S$_{2cm}$
	& S$_{7mm}$ & S$_{7mm}$ & Spectral \\
Source     & (h~m~s)~~~(\dec{~}{~}{~}) &(mJy) &($\frac{mJy}{beam}$) & (mJy)
        &($\frac{mJy}{beam}$) & (mJy) &($\frac{mJy}{beam}$)  
	& Index$^\dagger$ \\ 
\hline
\hline
VLA~1 (Ba) & 20 38 36.455 +42 37 34.80 & 5.3 & 3.5 & 4.0 & 1.5 
	& {\nodata} & {\nodata} & $~~0.2 \pm 0.3$\\
VLA~2      & 20 38 36.491 +42 37 34.30 & {\nodata} & {\nodata} 
	& 1.5 & 1.2 & 2.6 & 2.2 & $~~0.4 \pm 0.1$ \\
VLA~3 (Bb) & 20 38 36.491 +42 37 33.50 & 2.7 & 2.7 & 5.8 & 4.5 
	& 5.7 & 5.4 & $~~0.5 \pm 0.3$  \\
Bc         & 20 38 36.527 +42 37 31.50 & 4.4 & 2.8 & 1.7 & 1.2 
	& {\nodata} & {\nodata} & -$0.3 \pm 0.6$ \\
W75 N (A)    & 20 38 37.780 +42 37 59.00 & 116.1 & 4.5 
	& $> 99.7$ & 0.8 & {\nodata} &{\nodata} &{\nodata}\\
\hline
\end{tabular}

\vspace{.1in}
~~{\small $^\dagger$~Spectral index derived from this data and 
data from Torrelles et al. (1997) and Hunter et al. 1994.} 
\end{table}
\normalsize

Continuum emission at 7~mm wavelength can be due to a mixture of warm
dust and ionized gas emission.  Figure 2 (and Table 2) shows that the
flux density of the UC HII regions at 7~mm is consistent with or lower
than what is expected for ionized gas (either optically thin emission,
$S_\nu \propto \nu^{-0.1}$, or moderately optically thick).  How much
7~mm continuum emission is expected to be due to warm dust?  Using the
3~mm flux densities of Shepherd (2001) and assuming the thermal dust
emission between 3 \& 7~mm has a spectral index of 2, we expect to
detect about 36~mJy of combined flux at 7~mm from thermal dust
emission near the UC HII regions.  Only 8.3 mJy is recovered
suggesting that at least 80\% of the thermal dust emission is being
resolved out.  In the millimeter cores MM~2, MM~3, \& MM~4, 6--9 mJy
is expected while none is detected, again consistent with the thermal
dust emission being resolved out by the $0.2''$ resolution.

Source Bc is detected at 6 and 2~cm with a spectral index of $\alpha = -0.3
\pm 0.6$.  In a 1.3~cm continuum image, Torrelles et al. (1997) found
that Bc was marginally recovered at the $3 \sigma$ level when a
Gaussian taper was applied to the data suggesting that the source was
being resolved out by their higher $0.1''$ resolution.  Thus, the
negative spectral index derived for Bc may be due to missing flux at
short wavelengths (higher resolution) rather than an intrinsic
property of the source.  Bc is not detected in 7~mm continuum emission
with $\sim 0.2''$ resolution.  Based on the peak flux density at 2~cm
with $0.4''$ resolution, the expected peak emission at 7~mm is $\sim
0.3$\mjyb\ which is just under $1 \sigma$ in the 7~mm image.  Thus,
the surface brightness of the emission in Bc is below our sensitivity
limit.

VLA~3(Bb) is detected at all observed wavelengths.  The source
is marginally resolved at 2~cm \& 7~mm and the spectral index between
6~cm and 7~mm, $\alpha = 0.5 \pm 0.3$, is consistent with a moderately
optically thick HII region.

VLA~2 is detected between 2~cm and 7~mm with a spectral index
$\alpha = 0.4 \pm 0.1$, consistent with a moderately optically thick
HII region.  At 6~cm wavelength, the $1.2''$ resolution was not
adequate to isolate VLA~2 from VLA~1(Ba) or VLA~3(Bb).

VLA~1(Ba) is detected at 6 and 2~cm.  The elongation of the 2~cm
emission along the \water\ maser axis is consistent with that found by
Torrelles et al. (1997).  The spectral index between 6 and 1.3~cm is
$\alpha = 0.2 \pm 0.3$.  Assuming constant $\alpha = +0.2$ between
1.3~cm and 7~mm, the expected total flux density at 7~mm due to
ionized gas is 8.9~mJy while the expected peak flux density should be
roughly one fourth the peak found at 2~cm, i.e., $\sim 0.4${\mjyb}
(assuming the peak emission scales with the area of the synthesized
beam).  Figure 1 shows that there are a few peaks at the $3 \sigma$
level near the position of VLA~1(Ba) however the surface brightness
sensitivity is not adequate to recover the source structure.

Continuum emission associated with the extended HII region W75~N~A is
detected at a wavelength of 6~cm.  At 2~cm, most of the flux has been
resolved out, leaving only low-level emission and a few $3 \sigma$
peaks.  Previous images of the ionized gas in 6~cm continuum emission
were not able to recover the complex structure of this extended source
due to limited $uv$ coverage (Haschick et al. 1981).

\subsection{SiO emission}

SiO(v=0, J=2--1) channel maps with $\sim 5''$ and $3''$ resolution are
shown in Fig. 3.  The higher resolution image resolves out 58\% of the
flux density suggesting that a significant fraction of the emission is
diffuse.  SiO(v=0, J=2--1) red-shifted (10 to 17.8\kms) and
blue-shifted (2.23 to 10\kms) emission are shown in Fig. 4.  A zeroth
moment image covering the full velocity range is shown in Fig. 5 along
with representative SiO spectra at different locations in the cloud.

Despite the strong detection of SiO(v=0, J=2--1) with the OVRO
interferometer, the SiO(v=0, J=1--0) line could not be imaged well at
a resolution of $\sim 0.4''$.  For low J-transition lines, assuming
local thermodynamic equilibrium, optically thin emission, and high
excitation temperatures (T$_{ex} \, \ge \, 50$\,K), the brightness
temperature of SiO(v=0, J=2--1) should be roughly four times that of
SiO(v=0, J=1--0) (Goldsmith 1972).  Assuming the emission arises in
the same region, then:\\
\begin{equation}
  \frac{S_\nu(1-0)}{S_\nu(2-1)} = 
	\frac{T_B(1-0)}{T_B(2-1)}~~\frac{\lambda^2(2-1)}{\lambda^2(1-0)} 
	 = 0.0625
\end{equation}
Thus, SiO(J=1--0) peak flux densities of the order of 40\mjyb\ should
be observed, assuming similar $uv$ coverage between the VLA and OVRO.
Unfortunately, the $uv$ coverage of the interferometers differed
significantly.  To compare the two images, a 10~k$\lambda$ hole was
cut out of the OVRO data to match the hole in the VLA $uv$ coverage.
This provided a peak flux density in the SiO(v=0, J=2--1) line of
about 240\mjyb\ at a resolution of $3'' \times 2.5''$.  Roughly one
third of the total flux was recovered in this image.  The VLA
continuum-subtracted data cube then had a Gaussian taper applied until
the resolution matched that of OVRO.  Although the taper down-weighted
a significant fraction of the VLA data, the resulting peak emission in
the SiO(v=0, J=1--0) line was $\sim 30$\mjyb\ which is close to what
is predicted based on equation 1 above.  This exercises demonstrates
that the SiO(v=0, J=1--0) emission in W75~N is relatively diffuse.  

SiO line widths range from about 3 -- 10\kms\ within the W75~N cloud
(see Fig. 5).  Assuming a rotational temperature of 50K, the total SiO
gas mass traced by the (v=0, J=2--1) line is $2 \times
10^{-6}$~M{\sun} while the average column density is $3.5 \times
10^{14}$~cm$^{-2}$.  The estimated total mass of the cloud core based
on submillimeter continuum and CS(J=7-6) observations is $\sim 1000 -
2000$~M\sun\ (Moore, Mountain, \& Yamashita 1991; Hunter et al. 1994)
which implies an SiO/[\h] ratio of $\sim 5-7 \times 10^{-11}$.
The measured, average SiO fractional abundance, [SiO]/[\h], in typical
low-mass dark clouds is between $10^{-11}$ and $10^{-12}$ (Ziurys,
Friberg, \& Irvine 1989) while the average ambient abundances in
clouds with active star formation range from $10^{-10}$ to $10^{-11}$
(Codella, Bachiller, \& Reipurth 1999).  In comparison, SiO abundance
in the quiescent ridge of Orion is less than $3 \times 10^{-10}$
(Blake et al. 1987).  The SiO abundance calculated for the W75~N cloud
is consistent with the typical abundance for ambient gas in both high
and low-mass star forming regions. 

Note, SiO abundance can be as high as $10^{-5} - 10^{-6}$ in {\it
high-velocity} gas ($> 20${\kms} from the systemic velocity of the
cloud) which is directly associated with high and low-mass molecular
outflows (see, e.g., Martin-Pintado, Bachiller, \& Fuente 1992;
Shepherd, Churchwell, \& Wilner 1997).  However, the average abundance
enhancement is generally several orders of magnitude less as discussed
above.

\section{DISCUSSION}

\subsection{SiO emission and the molecular outflows}

Based on a comparison between CO(J=1--0), \h, and [FeII] emission,
STS03 suggested that only slow, non-dissociative J-type shocks exist
throughout the parsec-scale outflows produced by the central stars in
the W75~N (B) UC HII regions.  Fast, dissociative shocks, common in
jet-driven low-mass outflows, appear to be absent in W75~N.  Thus, the
energetics suggest that the outflows from the mid- to early-B
protostars in W75~N are not simply scaled-up versions of low-mass
outflows.  Further, there was no evidence for well-collimated,
parsec-scale jets such as those seen in flows from lower mass
protostars.  However, the observations of STS03 could not rule out the
presence of an underlying neutral jet that could drive the
CO outflows.

SiO emission in molecular flows is excited in shocks where silicon is
first removed from dust grains and then reacts with OH radicals to
form SiO in the post-shock cooling zone.  The gas phase abundance of
SiO can increase up to a factor of $10^6$ over that found in quiescent
molecular clouds and can delineate the axis of highly collimated
jet-driven outflows and/or the bow-shock where the head of a jet
interacts with dense molecular material (e.g. Haschick \& Ho 1990).
SiO 'jets' have been detected in well-collimated outflows from
low-mass YSOs (e.g. L1448: Guilloteau et al. 1992; HH~211: Chandler \&
Richer 2001) and in at least one massive outflow from an early-B star
(IRAS 20126+4014: Cesaroni et al. 1997, 1999).  Thus, SiO has the
potential to uncover collimated, molecular jets that may not be
obvious in other tracers.

Our SiO observations cover the central $60''$ field of the molecular
outflows mapped by STS03 (the full $5' \times 1.5'$ mosaic was not
covered).  There is physically diffuse SiO(J=2--1) and SiO(J=1--0)
emission centered near the positions of the UC HII regions.  The SiO
abundance is roughly a factor of 10 higher than abundances toward
dark, quiescent clouds and is consistent with what is expected for
ambient gas in in both low- and high-mass star forming regions.
Figure 4 shows the relation between the SiO emission, the locations of
the UC HII regions and millimeter cores 2--4, and the proposed
outflows from VLA~1~(Ba), VLA~3~(Bb), and MM~2 suggested by STS03.
There is no clear relationship between the SiO distribution and the
proposed outflows.  The higher resolution ($3''$) SiO images (Fig. 3
and right panel of Fig. 4) were made in an attempt to resolve out the
extended emission associated with the ambient gas and search for
compact, high-velocity, red- and blue-shifted emission that may
delineate collimated outflows.  There is no clear indication of a
jet-like structure from any specific source despite the presence of
VLA~1~(Ba), an HII region that appears to be excited by a thermal jet.

On a larger scale than was observed in SiO in this work, STS03 found
clear evidence for shocked gas associated with the outflows as seen
from the {\h} line morphology.  The shocked gas is diffuse and appears
to be caused by interactions between the ambient medium, wide-angle
outflowing gas, and ionized gas.  Our SiO observations of the center
$60''$ near the outflow driving sources show FWHM line widths up to
10\kms\ (Fig. 5) which suggests that the SiO is produced in shocks.
However, the resolution and sensitivity are not sufficient to
determine if the SiO abundance enhancement is associated with the CO
flows or if it arises from a shocked boundary between the ambient
medium and the ionized wind from the UC HII regions.

\subsection{Physical properties of the HII regions}

Three of four UC HII regions in W75~N (B) (VLA~1 (Ba), VLA~2, \& VLA~3
(Bb)) display evidence for on-going outflow/accretion based on
high-velocity molecular gas traced to the source and/or the presence
of {\water} or OH masers (Baart et al. 1986, Hunter et al. 1994,
Torrelles et al. 1997, STS03, Torrelles et al. 2003).  Thus, it is
likely that the ionized gas produced by the central star can escape
along the outflow axis.  The remaining two HII regions (Bc in W75~N
(B) \& W75~N (A)) are more extended structures that are recovered in
$\sim 1.2''$ resolution images at 6~cm but are resolved out at 2~cm
with higher resolution ($\sim 0.4''$).  As discussed in Wood \&
Churchwell (1989, hereafter WC89), complicated geometries present
difficulties for interpretation because physical parameters such as
density and surface brightness along a line of sight depend on source
structure.  Following the method outlined by WC89, we use the
integrated flux densities when knowledge of the source structure is
not required and, when geometry is important, we estimate peak values
using the peak flux densities per beam.  Table 3 presents the derived
physical parameters of the HII regions in W75~N.  For each source, the
values listed are: $\nu$, the frequency at which the derivations were
made; $\Delta s$, line-of-sight depth at the peak position (taken to
be the projected diameter of a sphere on the sky); T$_b$, the
synthesized beam brightness temperature; $\tau_\nu$, the peak optical
depth assuming the beam is uniformly filled with T$_e = 10^4$ K
ionized gas; EM, the emission measure in units of $10^7$ pc cm$^{-6}$;
$n_e$, the RMS electron density in units of $10^4$ cm$^{-3}$; U, the
excitation parameter of the ionized gas; $N_L$, the number of Lyman
continuum photons required to produce the observed emission assuming
an ionization-bounded, spherically symmetric, homogeneous HII region;
and finally, the spectral type of the central star assuming a single
ZAMS star is producing the observed Lyman continuum flux (Panagia
1973, WC89).

\small
\begin{table}[h]
\caption[]{Derived Parameters for HII regions} 

\smallskip
\begin{tabular}{|lcccccccccc|}
\hline
\multicolumn{3}{|c}{}    
 & \multicolumn{4}{c}{Peak Values from Observed } & 
 & \multicolumn{3}{c|}{Integrated Values from } \\
\multicolumn{3}{|c}{} 
 & \multicolumn{4}{c}{Flux Density per Synthesized Beam} & 
 & \multicolumn{3}{c|}{Integrated Flux Density} \\
\cline{4-7}\cline{9-11}
          & $\nu$ & $\Delta s$ & T$_b$ & $\tau_\nu$ & EM/10$^7$  
	    & $n_e/10^4$ & & U         & Log$N_L$   & Spectral \\ 
Source    & (GHz)& (pc)        & (K)        & 	    & (pc cm$^{-6}$)
          & (cm$^{-3}$)& & (pc cm$^{-2}$)& ($s^{-1}$)& Type \\ 

\hline
\hline
VLA~1 (Ba)$^\dagger$ &14.96 &0.006 &~52 &0.005 &1.5~ &5.0 & &3.2 &45.01 &B1 \\
VLA~2      &14.96 &0.004 &~42 &0.004 &0.38 &3.2 & &2.3 &44.59 &B2 \\
VLA~3 (Bb) &14.96 &0.007 &159 &0.016 &1.1~ &4.1 & &3.6 &45.17 &B1 \\
Bc         &~4.88 &0.024 &108 &0.011 &0.09 &0.6 & &3.2 &45.00 &B1.5 \\
W75 N (A)    &~4.88 &0.121 &172 &0.017 &0.15 &0.4 & &9.9 &46.42 &B0.5 \\
\hline
\end{tabular}

\vspace{.1in}
~~{\small $^\dagger$~Should be considered an upper limit due to likely
strong contamination by ionizing flux produced by shock waves in the
jet (see text for discussion). }
\end{table}
\normalsize

There are a number of errors associated with the values in Table 3
that are difficult to estimate due to our limited knowledge of the
detailed source structure.  First, peak properties in Table 1 assume
the beam is uniformly filled with $10^4$\,K gas.  However, the UC HII
regions are unresolved, thus peak properties should be considered
lower limits due to beam dilution effects.  Second, derivations based
on the integrated flux density should be considered lower limits for
sources with known outflows since the ionized gas can escape along the
outflow axis.  Third, there is no correction for dust absorption
within the ionized gas, which would tend to underestimate $N_L$, and
hence the spectral type of the star. Fourth, shock waves within the
outflow are expected to contribute to the total ionizing flux, which
would result in an overestimate of $N_L$.  Finally, accretion rates
above $10^{-5}$ to $10^{-4}$~M$_{\odot}$~yr$^{-1}$, typical for
early-B protostars, may inhibit the formation of a UC HII region near
the equatorial plane where accretion is highest (see, e.g., Churchwell
1999 and references therein).  Despite these uncertainties, the
derivations are probably accurate to within a spectral type, except,
perhaps, for VLA~1 (Ba).  Due to the elongation of the ionized gas
along the molecular outflow axis and the presence of {\water} masers
along the axis, Torrelles et al. (1997) \& STS03 argue that VLA~1 (Ba)
is a thermal jet source.  Thus, a significant fraction of the observed
centimeter continuum emission is likely due to the ionized jet rather
than emission from an ionization-bounded UC HII region produced by a
central, massive star.  If this is true, then the estimated spectral
type of VLA~1 (Ba) should be considered an upper limit.

For the sources VLA\,2 and VLA\,3 (Bb), our estimates of the physical
properties of the UC HII regions are lower than those of Torrelles et
al. (1997) for two reasons: 1) our estimates based on peak emission
likely suffer from beam dilution; and 2) the spectral indicies we
estimate from fits of the SEDs suggest optically thin or slightly
optically thick emission while Torrelles et al. estimated that the
emission was significantly optically thick (based on only two data
points).  Spectral types from both estimates still only differ by a
spectral type.  

Comparison of the values in Table 3 with those in WC89 (their Table
17), shows that the physical parameters of the ionized gas in the
W75~N sources are consistent with ZAMS stars with spectral types later
than B0.  Peak values of the emission measure and $n_e$ in the more
extended sources Bc \& W75~N (A) are roughly an order of magnitude
less than what is found in the more compact HII regions.  In the
absence of confinement, the radius of an HII region is expected to
increase with time as the ionization front expands to form an
increasingly larger Str\"{o}mgren sphere with subsequently smaller
electron densities.  Thus, the lower peak values and increased size of
the HII region are consistent with these sources being more evolved
than the more compact UC HII regions in W75~N (B).  Bc and W75~N (A)
also show no evidence for driving an outflow; again consistent with
the sources being more evolved.

STS03 assumed all millimeter continuum emission from W75~N (A) was due
to thermal dust and they calculated a mass of gas and dust to be
68~M\sun.  With a good centimeter continuum image, it is now possible
to estimate the likely contribution due to ionized gas and obtain a
better estimate for the molecular cloud mass traced by warm dust emission
surrounding the Str\"{o}mgren sphere.  Assuming the 6~cm emission from
W75~N (A) is optically thin ($S_\nu \propto \nu^{-0.1}$), the expected
flux density at 2.7~mm due to ionized gas is 85~mJy.  STS03 measured a
total flux density at 2.7~mm of 129.5~mJy, thus the expected mass in
the dust shell surrounding W75~N (A) is $\sim 23$~M\sun\ (see STS03
for a discussion of the assumptions and errors associated with this
estimate).

\subsection{Timescale for the formation of early-B stars in W75~N}

The timescale for the formation of O star clusters for which $M_\star
> 25$~M\sun\ appears to be less than 3~Myrs (Massy, Johnson, \&
DeGioia-Eastwood 1995).  In all the clusters studied by Massy et al.,
there was evidence for the continued formation of mid to early-B stars
(5-10~M\sun) at least 1~Myrs after the formation of the O stars.  Do
clusters where the most massive members are early-B stars have a
similar age spread as that found for clusters forming stars more
massive than 25~M\sun?

Within a radius of $30''$ ($\sim 0.3$ pc) from the W75~N (B) UC HII
regions are three young early-B stars: the central star in W75~N (A),
IRS1 and IRS2 (see, e.g., Fig. 10 of STS03).  The early-B stars are
embedded in a $\sim 1000 - 2000$~M\sun\ molecular cloud from which
multiple flows are emerging with a combined outflow mass of at least
250~M\sun\ (Moore et al. 1991, Hunter et al. 1994, STS03).  The
outflows are driven by at least two of the stars embedded in UC HII
regions and one millimeter core (MM~2).  No high velocity molecular
gas can be traced to the early B stars which are seen in the infrared
(W75~N (A), IRS1, and IRS2) which suggests that the infrared stars are
older than the central stars of the UC HII regions.  From the size and
velocity of the CO outflows, STS03 estimate that the central stars of
the UC HII regions are $\sim 10^5$ years old.

W75~N (A), a classic example of an expanding Str\"{o}mgren sphere
(Fig. 6), appears to be the oldest B-type star of the cluster.  The
central B0.5 star is detected in the near-infrared with an irregular
reflection nebula surrounding the star.  A $12''$ diameter sphere of
ionized gas surrounds the star (0.12 pc at a distance of 2 kpc) which,
in turn, is enclosed in an $18''$ (0.17 pc) diameter shell of
molecular gas.  The near-infrared colors are consistent with
foreground extinction from the molecular shell (A$_V \sim 20$, STS03).
IRS1 and IRS2, on the other hand, have excess emission at 2$\mu$m
suggesting that they have not had time to disperse their circumstellar
material via photoevaporation and stellar winds.  Thus, IRS1 \& IRS2
appear to be of an intermediate age between W75~N (A) and the embedded
stars in the UC HII regions.

Using the 6~cm observations of W75~N (A), we derive the age of
W75~N (A) and hence an estimate of the age spread of the
early-B stars in W75~N.  The age of W75~N (A) can be derived in two
ways: 1) from an estimate of the expansion time required for the HII
region to reach its current radius; and 2) assuming the central star
formed via accretion, from an estimate of the time it would take for
the remnant accretion disk to be photo-evaporated.  For the first case,
the expansion of an ionization-front at the boundary of an expanding
Str\"{o}mgren sphere increases with time from some initial radius
which depends on the ionizing flux from the central star and the
density of the ambient medium (Dyson \& Williams 1980).  Assuming a
strong shock approximation at the boundary of the ionization front,
the initial radius, $r_i$, is given by:
\begin{equation}
r_i = \left( \frac{3 N_L}{4 \pi~ n_o^2~ \beta_2} \right)^{\frac{1}{3}}
\end{equation}
and the time required for the HII region to expand to a radius $r(t)$
in a uniform density environment is: 
\begin{equation}
\tau = \left( \left[ \frac{r(t)}{r_i}\right]^{\frac{7}{4}} - 1 \right)
\left( \frac{4 r_i}{7 c_i}\right)
\end{equation}
where\\
\begin{tabular}{lll}
~~~ & $N_L$ & = the flux of ionizing photons \\
    & $n_o$ & = density of the ambient medium \\
    & $\beta_2$ & = recombination coefficient to all levels except the
	ground state at temperature, $T_e$ \\
    & $c_i$ & = sound speed in ionized gas ($\sim 10$\kms) \\
\end{tabular}
\\ 
Assuming $n_o = 2 \times 10^7$~cm$^{-3}$ (DePree, Rogr\'{\i}guez,
\& Goss 1995), $T_e = 10^4$~K, $\beta_2 = 2.6 \times 10^{-10}~
T_e^{-3/4}$~cm$^3$~s$^{-1}$, and $N_L = 2.6 \times 10^{46}$~s$^{-1}$
for this B0.5 star, then the initial radius of the Str\"{o}mgren
sphere is $r_i = 4 \times 10^9$~km and the expansion timescale for the
W75~N (A) HII region is $\tau = 1.2 \times 10^6$ years.

As discussed in DePree, Rogr\'{\i}guez, \& Goss (1995), this
calculation assumes that the molecular gas has a constant density and
infinite extent.  If this were the case, then one would expect W75~N (A)
to reach pressure equilibrium at a radius of only 0.01~pc,
which does not match the observed radius of $> 0.1$ pc.  Instead,
we expect the molecular gas density to decrease with radius which
suggests that, for timescales $\gtsim\ 10^5$ years, UC HII regions are
probably not in pressure equilibrium and should be expanding.  At the
same time, the molecular material surrounding the Str\"{o}mgren sphere
is expected to expand at a slower rate ($\sim 1$\kms) due to the lower
temperature and molecular composition.  Thus, the exact expansion
timescale of the HII region and the stability of the molecular
envelope surrounding the ionized gas are quite uncertain.  

For the second method, we assume the central star formed via
accretion.  Although competing theories exist for how the most massive
O stars formed (e.g. coalescence or accretion), observational evidence
for disks around early-B stars and the similarity between outflows
produced by low-mass stars and those from mid- to early-B stars,
suggest that stars up to spectral type B0 or O9 most likely form via
accretion (see, e.g., the review by Shepherd 2003 and references
therein).  If the central star of W75~N (A) once had a massive
accretion disk, then the lifetime of the UC HII region could be
lengthened due to the photoevaporation of the circumstellar disk by
the stellar wind (Hollenbach et al. 1994).  The disk material would
have provided high-density, ionized gas thus, the UC HII region would
persist as long as the disk survived the mass loss (assuming the disk
is no longer being fed by material from the surrounding molecular
core).  Since the near-infrared colors suggest there is no current
disk, we assume the disk has been completely photo-evaporated and the
timescale for this to occur would represent a lower limit to the age
of W75~N (A).  For the ``weak wind'' case of Hollenbach et
al. appropriate for early-B stars, the lifetime of the disk is given
by:
\begin{equation}
\tau_{disk} = 7 \times 10^4 ~\Phi_{49}^{-1/2} ~M_1^{-1/2}~M_d ~~{\rm [yrs]}
\end{equation}
where\\
\begin{tabular}{lll}
~~~ & $\Phi_{49}$ & = ionizing Lyman continuum flux in units of
    $10^{49}$~s$^{-1}$  \\
    & $M_1$ & = the mass of the central star in units of 10~M\sun \\
    & $M_d$ & = disk mass in units of M\sun \\
\end{tabular}
\\ 
For W75~N (A), $\Phi_{49} = 2.6 \times 10^{-3}$ and $M_1 \sim 1.5$.
Shu et al. (1990) showed that an accretion disk becomes
gravitationally unstable when it reaches a mass of $M_d \sim 0.3
M_\star$ where $M_\star$ is the mass of the central protostar.  During
the initial collapse of the cloud core, the disk mass may be
maintained close to the value of $0.3 M_\star$.  When infall ceases
and the disk mass falls below the critical value, disk accretion onto
the star may rapidly decline and photoevaporation may be the dominant
mechanism which disperses the remaining gas and dust (Hollenbach et
a. 1994).  Based on this scenario, we assume an initial
disk at the edge of stability, that is $M_d \sim 0.3 M_\star = 4.5
M$\sun.  Errors in the estimate for the photoevaporative timescale
would scale directly as $M_d$.  We find that $\tau_{disk} = 5 \times
10^6$ years.  

Both estimates for the lifetime of W75~N (A) have numerous assumptions
about, e.g. characteristic cloud densities, disk mass, \& temperature
of the ionized gas.  None-the-less, they are probably reasonable to
within an order of magnitude.  These derivations suggest that the B0.5
star in W75~N (A) is roughly $1-5 \times 10^6$ years old while the
youngest B stars forming are $\sim 10^5$ years old.  Thus, the spread
in ages between young B-stars in this cluster is $\Delta\tau = 0.1-5
\times 10^6$ years.

Efremov \& Elmegreen (1998) and Elmegreen et al. (2000) suggest that
the duration of star formation tends to vary with the size, $S$, of the
cluster as something like the crossing time for turbulent motions,
e.g. $\Delta\tau \propto S^{0.5}$.  Given a cluster diameter of W75~N
of $\sim 1'$ (0.6 pc), the expected age would be roughly 0.8 Myrs,
which is somewhat lower than our estimates although still within the
errors.  Comparing with other observations, clusters forming stars
with $M_\star > 25$~M\sun\ have a typical spread in ages,
$\Delta\tau$, of about 2~Myrs for the O stars while mid- to early-B
stars continue to form for at least another million years (Massy,
Johnson, \& DeGioia-Eastwood 1995).  Thus, the early-B stars in W75~N
are formed over a period that is consistent with the timescale for
early-B stars formed in clusters with more massive stars.

\section{SUMMARY}

We have observed the W75~N massive star forming region in SiO(J=2--1)
\& (J=1--0) to search for well-collimated neutral jets from the
early-B protostars and in centimeter and 7~mm continuum emission to
examine the nature of the driving sources.  The SiO emission is
diffuse with no clear indication of a neutral, collimated jet from the
region.  This does not, however, completely rule out the presence of a
jet since enhanced SiO emission does not always trace known jet
structure.

The ionized gas surrounding the protostars have emission measures,
densities, and derived spectral types which are consistent with
early-B stars.  Most of the detected sources have spectral indicies
which suggest optically thin to moderately optically thick HII regions
produced by a central ionizing star.

By comparing the oldest and youngest B stars in the cluster, an
estimate for the duration of early-B star formation in W75~N is
obtained.  The oldest star in the cluster is roughly $1-5 \times 10^6$
years old while the youngest, B protostars are $\sim 10^5$ years old.
Thus, the spread in ages is $0.1-5 \times 10^6$ years.  The
age spread for W75~N is consistent with that found for early-B stars
born in clusters forming more massive stars ($M_\star > 25$M\sun).

\vspace{-4mm} \acknowledgements 
{\bf Acknowledgments:} Research at the Owens Valley Radio Observatory
is supported by the National Science Foundation through NSF grant
number AST 96-13717.  Star formation research at Owens Valley is also
support by NASA's Origins of Solar Systems program, grant NAGW-4030
and by the Norris Planetary Origins Project.  S. Kurtz acknowledges
support from Project IN118401 and CONACyT Project E-36568.
D. Shepherd would like to thank Sally Oey for useful discussions on
the evolutionary timescales of young clusters.

\clearpage
\begin{center}  {\bf Figure Captions}  \end{center}

\noindent
{\bf Figure~1.~~} 
Continuum emission from W75~N~(A) and UC HII regions within W75~N~(B)
at 6~cm (top), 2~cm (lower left), and 7~mm (lower right) wavelength is
plotted in both contours and greyscale.
{\bf Top:} The 6~cm image RMS is 0.11~\mjyb; contours are plotted at
--3, 3, 5~$\sigma$ and continue with spacings of 5~$\sigma$; greyscale
is displayed on a linear scale from 0.33~\mjyb\ to 4.5~\mjyb.  The
synthesized beam, plotted in the lower left corner, is $1.36'' \times
1.12''$ at position angle 29.2\deg.  The extended emission from
W75~N~(A) coincides with the millimeter core MM~5 and a near-infrared
reflection nebula and star (Shepherd, Testi, \& Stark 2003).
Locations of UC HII regions in W75~N~(B) that are detected in 6~cm
continuum emission are identified by filled triangles.  The dashed box
around the W75~N~B sources shows the area displayed in the lower
two images at 2~cm and 7~mm.  
{\bf Lower left:} The 2~cm image RMS is 0.23~\mjyb; contours are
plotted at --3, 3, 4, 5, 6, 8,~$\sigma$ and continue with spacings of
2~$\sigma$; greyscale is displayed from 0.46~\mjyb\ to 4.42~{\mjyb}.
The synthesized beam, plotted in the lower right corner, is $0.44''
\times 0.36''$ at P.A. 31.4\deg.
%
{\bf Lower right:} The 7~mm image RMS is 0.31~\mjyb; contours are
plotted at --3, 3, 5, 7~$\sigma$ and continue with spacings of
2~$\sigma$; greyscale is displayed from 0.62~\mjyb\ to 5.4~\mjyb.  The
synthesized beam, plotted in the lower right corner, is $0.27'' \times
0.20''$ at P.A. 89.4\deg.  UC HII region Bc is not detected.  The
locations of the UC HII regions in the field are indicated with filled
triangles.  Water masers are shown as crosses (Torrelles et al. 1997);
OH maser positions are represented as filled circles (Baart et
al. 1986).

\noindent
{\bf Figure~2.~~}
The spectral energy distributions of the UC HII regions in W75~N~(B).  
Asterisks represent data from this work, squares represent data
from Hunter et al. (1994), circles are from Torrelles et al. (1997),
and the triangle represents the $0.9''$ resolution 1~mm data from
Shepherd (2001).  Upper limits ($3 \sigma$) are shown as symbols with
arrows.  In all cases, estimated errors are smaller than the symbols.  
Solid lines show linear least squares fits to the SEDs for data between
6~cm and 7~mm.  The slope of the fit (spectral index) is shown in the
lower right corner of each plot.  

\noindent
{\bf Figure 3.~~}
{\bf Top:} SiO(v=0, J=2--1) channel maps at 1.726\kms\ spectral
resolution between 0.5 and $19.5$\kms.  The velocity is indicated in
the upper right of each panel.  The RMS is 45\mjyb.  Contours are
plotted from $\pm 3, 4, 5, 6~ \sigma$ and continue with a spacing of
$2~ \sigma$.  The last panel (velocity 19.5\kms) shows the synthesized
beam in the lower right corner ($5.35'' \times 4.15''$ at
P.A. $-60.3^\circ$) and a scale size of 0.1 pc is represented by a bar
in the lower left corner.  The locations of four millimeter continuum
peaks in W75~N~(B) are shown as plus signs (MM~1 -- MM~4, Shepherd
2001) while positions for the UC HII regions VLA 1(Ba), VLA2, VLA
3(Bb), and Bc are indicated by small triangles (Hunter et al. 1994;
Torrelles et al. 1997).
{\bf Bottom:} SiO(v=0, J=2--1) channel maps made with a higher spatial
resolution of $3.17'' \times 2.56''$ at P.A. $-61.5^\circ$ (beam shown
in lower right panel).  The RMS is 39\mjyb.  Contours are plotted from
$\pm 3, 4, 5, 6~ \sigma$ and continue with a spacing of $2~ \sigma$.

\noindent
{\bf Figure 4.~~}
Blue-shifted (2.23 to 10\kms; thin lines) and red-shifted (10 to
17.8\kms; thick lines) SiO(v=0, J=2--1) emission contours.  The left
panel shows the lower resolution data while the right panel shows the
higher resolution data.  The synthesized beam is shown in the lower
right corner of each image.  Millimeter cores MM~2, MM~3, and MM~4 are
shown as filled circles, UC HII regions within MM~1 are indicated by
filled triangles.  Position angles of outflows proposed by Shepherd,
Testi, \& Stark (2003) are illustrated by arrows.

\noindent
{\bf Figure~5.~~}
W75~N SiO(v=0, J=2--1) zeroth moment map and SiO spectra.  The
SiO(v=0) image (lower left) has an RMS of 0.3{\mjybkms}; contours
begin at $\pm 3, 4, 6 \sigma$ and continue with spacings of $2
\sigma$.  The synthesized beam of $5.35'' \times 4.15''$ at
P.A. $-60.3^\circ$ is shown in the lower right corner while a scale
size of 0.1~pc is shown in the lower left corner.  Millimeter cores
MM~2, MM~3, and MM~4 are shown as filled circles, UC HII regions
within MM~1 are indicated by filled triangles.  SiO(v=0, J=2--1)
spectra are shown at different locations in the cloud.  The dashed
vertical line in each plot represents $v_{LSR} = 10$\kms.

\noindent
{\bf Figure~6.~~} 
W75~N (A):  White contours represent the 6~cm continuum emission
tracing ionized gas in the Str\"{o}mgren sphere surrounding the
central B0.5 star (contours are the same as in Fig. 1). Greyscale
shows the near-infrared K-band reflection nebula and central star
while the black contours represent the 3~mm continuum emission from 
the warm dust shell surrounding the ionized gas (from Fig. 5 of
STS03).


\end{document}